\documentclass[reprint,showpacs,preprintnumbers,amsmath,amssymb,aps,superscriptaddress,aas_macros,nofootinbib]{revtex4-1}
\usepackage[dvipdfmx]{graphicx}
\usepackage{dcolumn}% Align table columns on decimal point
\usepackage{bm}% bold math
\usepackage{aas_macros,braket}
\input{colordvi.tex}
\usepackage{ulem}
\usepackage{color}
\usepackage{hyperref}

\begin{document}
%\draft
\preprint{}
\title{Thermal Sunyaev-Zel'dovich effect in the intergalactic medium with primordial magnetic fields}
\author{Teppei Minoda}
\email{minoda.teppei@d.mbox.nagoya-u.ac.jp}
\affiliation{%
Department of Physics and Astrophysics, Nagoya University, Nagoya 464-8602, Japan
}
\author{Kenji Hasegawa}
\email{hasegawa.kenji@a.mbox.nagoya-u.ac.jp}
\affiliation{%
Department of Physics and Astrophysics, Nagoya University, Nagoya 464-8602, Japan
}
\author{Hiroyuki Tashiro} 
\email{hiroyuki.tashiro@nagoya-u.jp}
\affiliation{%
Department of Physics and Astrophysics, Nagoya University, Nagoya 464-8602, Japan
}
\author{Kiyotomo Ichiki}
\email{ichiki@a.phys.nagoya-u.ac.jp}
\affiliation{%
Department of Physics and Astrophysics, Nagoya University, Nagoya 464-8602, Japan
}
\affiliation{%
Kobayashi-Maskawa Institute for the Origin of Particles and the Universe, Nagoya University,
Chikusa-ku, Nagoya 464-8602, Japan
}
\author{Naoshi Sugiyama}
\email{naoshi@nagoya-u.jp}
\affiliation{%
Department of Physics and Astrophysics, Nagoya University, Nagoya 464-8602, Japan
}
\affiliation{%
Kobayashi-Maskawa Institute for the Origin of Particles and
the Universe, Nagoya University, Chikusa-ku, Nagoya 464-8602, Japan
}
\affiliation{%
Kavli Institute for the Physics and Mathematics of the Universe (Kavli IPMU),
The University of Tokyo, Chiba 277-8582, Japan
}

\date{\today}
\begin{abstract}
The presence of ubiquitous magnetic fields in the Universe is suggested
from observations of radiation and cosmic ray from galaxies or the intergalactic medium~(IGM).
One possible origin of cosmic magnetic fields is the magnetogenesis in the primordial Universe.
Such magnetic fields are called primordial magnetic fields (PMFs),
and are considered to affect the evolution of matter density fluctuations and the thermal history of the IGM gas.
Hence the information of PMFs is expected to be imprinted on the anisotropies
of the cosmic microwave background (CMB)
through the thermal Sunyaev-Zel'dovich (tSZ) effect in the IGM.
In this study, given an initial power spectrum of PMFs as
$P(k)\propto B_{\rm 1Mpc}^2 k^{n_{B}}$,
we calculate dynamical and thermal evolutions of the IGM under the influence of PMFs,
and compute the resultant angular power spectrum of the Compton $y$-parameter on the sky.
As a result, we find that two physical processes driven by PMFs
dominantly determine the power spectrum of the Compton $y$-parameter;
(i) the heating due to the ambipolar diffusion effectively works
to increase the temperature and the ionization fraction,
and (ii) the Lorentz force drastically enhances the density contrast on small scale
just after the recombination epoch. 
These facts result in making the anisotropies of the CMB temperature on small scales,
and we find that the signal goes up to $10~\mu \rm{K}^2$ around $\ell \sim 10^6$
with $B_{\rm 1~Mpc}=0.1$~nG and $n_{B}=0.0$.
Therefore, CMB measurements on such small scales may provide a hint for the existence of the PMFs.
 \end{abstract}
\maketitle

\section{Introduction}
Magnetic fields are found on various scales in the Universe
through numerous astronomical observations so far.
They appear to be not only on small scales
such as in planets, stars \cite{1998FCPh...19..319V}, and galaxies \cite{1996Natur.379...47B},
but also in galaxies at high redshifts~\cite{2008Natur.454..302B},
and even on large scales such as in galaxy clusters~\cite{2002ARA&A..40..319C}.
Moreover, there are some works which suggest the presence of magnetic fields in the intergalactic region.
They have provided lower limits on the strength of intergalactic magnetic fields
from observations of $\gamma$-rays emitted from distant blazars
~\cite{2010Sci...328...73N, 2013ApJ...771L..42T,
2010ApJ...722L..39A, 2015PhRvL.115u1103C, 2011APh....35..135E}.
They claim that magnetic fields on kpc scales should be stronger than $3\times10^{-16}$ G in void regions.
Interestingly, this lower limit seems to be nearly the same among these studies
though they used different methods and observational data.

It is still ambiguous that the origin of these magnetic fields is either
an astrophysical or a cosmological phenomenon.  If magnetic fields
existed in a protogalaxy, it is believed that the magnetic fields would
be amplified by the galactic dynamo process
\cite{1970ApJ...160..383P,2005PhR...417....1B}.
According to the recent study about the cosmological dynamo process,
seed fields as large as $10^{-20}$--$10^{-30}$ G are enough
to explain the galactic magnetic fields of a few $\mu$G
observed in the present Universe \cite{1999PhRvD..60b1301D}.
However, the efficiency of the dynamo process is still under discussion.
A robust way of magnetic field amplification in the cosmological context
is the adiabatic amplification during the structure formation.
In this case, nG magnetic fields are required for $\mu$G magnetic fields
observed in galaxies and galaxy clusters.
Many mechanisms have been proposed to explain the generation of such seed fields
with both astrophysical~\cite{2008RPPh...71d6901K}
and cosmological phenomena~\cite{1988PhRvD..37.2743T}. 
In particular, magnetic fields generated by cosmological phenomena
seem to have a larger correlation length,
and they are called primordial magnetic fields~(PMFs).
There are many scenarios of the PMF generation in various epochs of the early universe
(for a recent review, see Ref.~\cite{2016RPPh...79g6901S}).

The properties of PMFs depend on the generation mechanisms,
and are often characterized by $B_\lambda$,
the field strength on a scale $\lambda$ at the present epoch.
Through cosmological observations,
many works have provided observational constraints on $B_\lambda$
~(for a review, see~\cite{2013A&ARv..21...62D}).
One of the important constraints comes from the anisotropies of the cosmic microwave background (CMB).
PMFs induce the magnetohydrodynamics~(MHD) motions in the primordial ionized plasma
by the Lorentz force on small scales.
Additionally, the stress-energy tensor of PMFs also generates large scale metric perturbations
\cite{2002PhRvD..65l3004M, 2004IJMPD..13..391G, 2010PhRvD..81d3517S}. 
As a result,
these fluctuations generate additional anisotropies in the CMB temperature and polarization
on both large and small scales~\cite{1998PhRvL..81.3575S, 2003MNRAS.344L..31S,
2008PhRvD..77f3003G, 2006PhRvD..74f3002G, 2009PhRvD..79j3007G, 2009PhRvD..79l1302G}.  
The Planck collaboration has performed a data analysis by taking these effects into account consistently.
The obtained constraint on the amplitude of PMFs is
$B_{1\rm{Mpc}} \lesssim 5$ nG considering the these effects~\cite{2016A&A...594A..19P}.

Even after the recombination epoch, PMFs may also provide significant effects in the expanding universe
In particular, PMFs induce density fluctuations by the Lorentz force
~\cite{1978ApJ...224..337W, 1996ApJ...468...28K},
and heat the baryon gas in the intergalactic medium~(IGM)
due to the dynamical friction between neutral and charged particles
that is called the ambipolar diffusion~\cite{1956MNRAS.116..114C}.
These effects give strong impacts on the structure formation, the reionization process,
and so on~\cite{2005MNRAS.356..778S,2006MNRAS.368..965T}.
Many authors have discussed the observational constraints on PMFs
from the Thomson optical depth
~\cite{2015JCAP...06..027K,2005MNRAS.356..778S,2006MNRAS.368..965T},
21~cm anisotropies~\cite{2006MNRAS.372.1060T, 2009JCAP...11..021S, 2014PhRvD..89j3522S},
cosmic shear~\cite{2012JCAP...11..055F},
and galaxy surveys~\cite{2012MNRAS.424..927T,2013ApJ...770...47K}.

We investigate the thermal Sunyaev-Zel'dovich~(tSZ) effect in the IGM due to PMFs.
We assume that PMFs are random Gaussian fields.
Therefore, the heating efficiency of the ambipolar diffusion is not spatially homogeneous.
As a result, the fluctuations of the IGM gas temperature arise.
Additionally, the random magnetic fields can also generate the matter density fluctuations.
These fluctuations create the anisotropies of the Compton $y$-parameter on the sky
and result in the CMB temperature anisotropies due to the tSZ effect.
The aim of this paper is to investigate the potential
to constrain the strength and scale dependence of PMFs
through measurements of the tSZ angular power spectrum.
It is worth mentioning the difference of this work from previous ones
~\cite{2011MNRAS.411.1284T,2012PhRvD..86d3510S}.
Although these works investigated the effect of PMFs on the tSZ angular power spectrum,
their concerns are in the tSZ effect in galaxy clusters.
The additional density fluctuations generated by PMFs increase the abundance of galaxy clusters
and enhance the tSZ angular power spectrum.
On the other hand, we focus on the tSZ effect in the IGM in the present study.
We evaluate the fluctuations of the density, temperature and ionization fraction
of the IGM gas with PMFs and,
then, compute the tSZ angular power spectrum.

In the evolutional equations for the IGM gas density and temperature,
the Lorentz force and the ambipolar diffusion are represented as
nonlinear convolutions of PMFs. Therefore, it is difficult to evaluate
the effects of PMFs on the fluctuation of the IGM gas density and
temperature analytically.  In this investigation, therefore, we perform
numerical simulations.  In a simulation box, we generate PMFs based on a
simple power-law model, assuming that PMFs are adiabatically decaying
due to the cosmological expansion.  We compute the gas density evolution
and the thermal history of the IGM gas with the ambipolar diffusion and
the several cooling effects of the IGM gas explained
in~\cite{1994MNRAS.269..563F}.
Although the linear perturbations are employed to calculate the density evolution for the simplicity, 
 this work is the first attempt to evaluate the PMF effect on the spatial distributions of the IGM density,
temperature and ionization fraction, consistently.  Finally we show the
resultant tSZ angular power spectrum and discuss its dependence on the
statistical properties of PMFs.

This paper is organized as follows:
In Sec.~$\rm{II}$, we describe how PMFs alter the dynamics and the thermal history of the IGM
and create tSZ effect anisotropies.
In Sec.~$\rm{III}$, we introduce our numerical method
to realize PMFs and to compute the thermal history and the tSZ angular power spectrum.
Section IV is dedicated for results and discussions.
Section V concludes this paper.
Throughout this paper
we adopt the flat-$\Lambda$CDM model with the cosmological parameters
from the {\it Planck}
2015 results \cite{2016A&A...594A..13P}:
$H_0=67.8$ km/s/Mpc, $\Omega_\Lambda=0.692$,
$\Omega_\mathrm{m}=0.308$ and $\Omega_\mathrm{b}=0.048$.
%%%%%%%%%%%%%%%%%%%%%%%%%%%%%%%%%%%%%%%%%%%%%%%%%%%%%%%%

\section{cosmic magnetism and SZ effect}
After the recombination epoch, the presence of magnetic fields could
change the gas dynamics.  In this study, we consider two effects: (A)
generation of the matter density fluctuations by PMFs, and (B) heating
of the IGM gas due to the ambipolar diffusion of PMFs.  Then, we
evaluate the tSZ effects taking these two effects into account
simultaneously.

\subsection{IGM density fluctuations due to PMFs}
As investigated in the previous
study~\cite{1978ApJ...224..337W,1996ApJ...468...28K}, magnetic fields
can generate baryon~(IGM) density fluctuations after the recombination
epoch.  The evolutional equations of the density fluctuations for cold
dark matter and baryons can be provided as~\cite{2005MNRAS.356..778S}
\begin{eqnarray}
&&\cfrac{\partial^2 \delta_\mathrm{c}}{\partial t^2} + 2 H(t) \cfrac{\partial \delta_\mathrm{c}}{\partial t}
- 4\pi G (\rho_\mathrm{c} \delta_\mathrm{c} + \rho_\mathrm{b} \delta_\mathrm{b}) = 0,
\label{eq:cdm}
\\
&&\cfrac{\partial^2 \delta_\mathrm{b}}{\partial t^2} + 2 H(t)\cfrac{\partial \delta_\mathrm{b}}{\partial t}
- 4\pi G (\rho_\mathrm{c} \delta_\mathrm{c} + \rho_\mathrm{b} \delta_\mathrm{b}) = S(t),
\label{eq:baryon}
\end{eqnarray}
where $H(t)$ represents the Hubble parameter, $\rho_\mathrm{b,c}$ and
$\delta_\mathrm{b,c}$ are the densities and density contrasts of baryons (b)
and cold dark matter (c), respectively.

In Eq.~(\ref{eq:baryon}), $S(t)$ is the source term due to the Lorentz
force of PMFs that is given by
\begin{equation}
S(t) = \cfrac{\nabla \cdot (\nabla \times \mathbf{B}(t,\mathbf{x}))
 \times \mathbf{B}(t,\mathbf{x})}{4 \pi \rho_\mathrm{b} (t) a^2(t)},
 \label{eq:source_b}
 \end{equation}
where $\mathbf{B}(t,\mathbf{x})$ represents the PMFs at a comoving three-dimensional position $\bf x$ with time $t$, $a(t)$ is the cosmic scale factor, and $\nabla$ is taken in the comoving coordinate.
We can obtain the solution of Eq.~(\ref{eq:baryon}) analytically by the Green function method.
Resultantly, assuming a matter dominated universe
and $\delta_\mathrm{b}=0$ initially, we can write the evolution of
$\delta _\mathrm{b}$ as
\begin{widetext}
\begin{eqnarray}
\delta_\mathrm{b} = \cfrac{2S(t)}{15H^{2}(t)}
  \left[ \left\{ 3\left(\cfrac{a}{a_\mathrm{rec}} \right)  \right. \right.
   &+& 2\left(\cfrac{a}{a_\mathrm{rec}} \right)^{-\frac{3}{2}}
  - \left. 15 \ln \left( \cfrac{a}{a_\mathrm{rec}}\right) \right\}
   \cfrac{\Omega_\mathrm{b}}{\Omega_\mathrm{m}}  \nonumber \\
 &+& 15 \ln \left( \cfrac{a}{a_\mathrm{rec}}\right)
 + 30 \left(1 - \cfrac{\Omega_\mathrm{b}}{\Omega_\mathrm{m}}\right)
  \left(\cfrac{a}{a_\mathrm{rec}}\right)^{-\frac{1}{2}}
   -  \left. \left(30-25\cfrac{\Omega_\mathrm{b}}{\Omega_\mathrm{m}}\right)  \right],
\label{eq:delta_b}
\end{eqnarray}
\end{widetext}
where $\Omega_\mathrm{m}$ and $\Omega_\mathrm{b}$
are the density parameters of total matter and baryons, respectively,
and $a_\mathrm{rec}$ is the cosmic scale factor at the recombination epoch.
Note that Eqs.~(\ref{eq:cdm}) and (\ref{eq:baryon}) are valid only when $\delta_\mathrm{b} \ll 1$
and the baryon pressure is negligible.
We discuss the validity of these assumptions later.

\subsection{Thermal history of the IGM gas}
The existence of PMFs significantly affects the thermal history of the
IGM gas, in addition to its density fluctuations.
After the photon decoupling, the cosmic ionization fraction rapidly decreases.
However there still exist residual ionized particles in the IGM.
These ionized particles are affected by the Lorentz force of the PMFs,
while neutral particles do not feel the force.
Therefore, a relative motion between ionized and neutral gases arises.
The energy of this relative motion is thermalized by subsequent collisions
between ionized and neutral particles.
This process is known as the ambipolar diffusion
and the magnetic fields provide the extra heating of the IGM temperature.
Through this process, the energy of PMFs dissipates and the thermal energy of the IGM gas increases.

The evolutional equation of the gas temperature $T_{\rm gas}$ with the extra heating
is given by~\cite{1994MNRAS.269..563F}
\begin{eqnarray}
\cfrac{d T_{\rm gas}}{dt} = &-& 2 H(t) T_{\rm gas} + \cfrac{\dot{\delta_\mathrm{b}}}{1+\delta_\mathrm{b}} T_{\rm gas} \nonumber \\
 &+& \cfrac{x_{\mathrm{i}}}{1+x_{\mathrm{i}}} \cfrac{8 \rho_\gamma \sigma_\mathrm{T}}{3 m_\mathrm{e} c} (T_{\gamma} -T_{\mathrm{gas}})
 + \cfrac{\Gamma(t)}{1.5 k_\mathrm{B} n_\mathrm{b}} \nonumber \\
 &-& \cfrac{x_{\mathrm{i}} n_\mathrm{b}}{1.5k_\mathrm{B}} [\Theta x_{\mathrm{i}} + \Psi(1-x_{\mathrm{i}}) + \eta x_{\mathrm{i}} + \zeta (1-x_{\mathrm{i}})], \nonumber \\
 \label{eq:th}
\end{eqnarray}
where $x_{\mathrm{i}}$ is the ionization fraction, $m_\mathrm{e}$ is the electron mass,
$\sigma_\mathrm{T}$ is the cross section of the Thomson scattering,
$k_{\rm B}$ is the Boltzmann constant, $n_\mathrm{b}$ is the baryon number density,
and the subscript $\gamma$ represents the
values for the CMB.  The first term in the right-hand side of equation
(\ref{eq:th}) represents the adiabatic cooling by the expansion of the
Universe, the second is the effect of adiabatic compression or expansion
due to the growth of the local density, and the third describes the
Compton cooling (or heating).  The forth term represents the extra
heating source of the IGM gas.  Here we consider the ambipolar diffusion
due to the PMF, whose heating rate $\Gamma(t)$ is given
by~\cite{2005MNRAS.356..778S}
\begin{equation}
\Gamma(t) = \frac{|(\nabla \times {\mathbf{B}}(t,\mathbf{x}))
 \times {\mathbf{B}}(t,\mathbf{x})|^2}{16\pi^2 \xi \rho_\mathrm{b}^2 (t)}
 \cfrac{(1-x_{\mathrm{i}})}{x_{\mathrm{i}}},
 \label{eq:source_t}
\end{equation}
where $\xi$ is the drag coefficient,
and $\xi = 3.5 \times 10^{13}~{\rm cm^3~g^{-1}~s^{-1}}$~\cite{1992pavi.book.....S}.
The last term interposed between the square brackets stands for the cooling effects of the gas.
Here we include the free-free cooling~(bremsstrahlung), the collisional excitation cooling,
the recombination cooling and the collisional ionization cooling.
In Eq.~\eqref{eq:th}, these cooling rates are denoted as $\Theta$, $\Psi$, $\eta$, and $\zeta$, respectively,
and we adopt the values in Ref.~\cite{1994MNRAS.269..563F}.

In order to solve the thermal history in Eq.~(\ref{eq:th}), the ionization fraction $x_{\mathrm{i}}$ is required.
According to Refs.~\cite{2005MNRAS.356..778S, 2000ApJS..128..407S},
the ionization history is given by
\begin{eqnarray}
 {d  x_{\mathrm{i}} \over dt}
= && \left[- \alpha_e n_\mathrm{b} x_{\mathrm{i}}^2 + \beta_e (1-x_{\mathrm{i}})
  \exp \left( \cfrac{E_{1s} - E_{2s}}{k_\mathrm{B} T_\gamma} \right ) \right] D \nonumber \\
 && + \gamma_e n_\mathrm{b} (1-x_{\mathrm{i}})x_{\mathrm{i}},
 \label{eq:ion}
\end{eqnarray}
where we adopt the three-levels model~(1s, 2s+2p and continuum),
$E_i$ represents the binding energy of $i$-state~($E_i<0$),
and $D$ is the suppression factor due to the Ly-$\alpha$ resonance photons,
\begin{equation}
D = \cfrac{1+K\Lambda n_\mathrm{b} (1-x_{\mathrm{i}})}
		{1+K\Lambda n_\mathrm{b} (1-x_{\mathrm{i}}) + K \beta_e (1-x_{\mathrm{i}})},
\end{equation}
with the redshift rate of Ly-$\alpha$ denoted by $K$,
and the two-photon emission coefficient denoted by
$\Lambda = 8.22458~\mathrm{s}^{-1}$~\cite{2000ApJS..128..407S}.
The first, second, and last terms in the right-hand side in Eq.~\eqref{eq:ion} represent
the collisional recombination, the photoionization and the collisional ionization, respectively.
The coefficients in those terms are given by
\begin{eqnarray}
\alpha_e &=& 1.14 \times 10^{-13} \times
 \cfrac{4.309~T_4^{-0.6166}}{1+0.6703~ T_{4}^{0.5300}} ~~ [\mathrm{cm}^3~\mathrm{s}^{-1}],
\\
\beta_e &=& \alpha_e
 \left(\cfrac{2 \pi m_\mathrm{e} k_\mathrm{B} T_\gamma}{h_\mathrm{Pl}^2} \right)^\frac{3}{2}
 \exp \left(\frac{E_{2s}}{k_\mathrm{B} T_\gamma} \right) ~~ [\mathrm{s}^{-1}],
 \label{beta_e} 
\\
 \gamma_e &=& 0.291 \times 10^{-7}\times  U^{0.39}\frac{\exp(-U)}{0.232+U}
  ~~ [\mathrm{cm}^3~\mathrm{s}^{-1}],
\end{eqnarray}
with $T_4=T_\mathrm{gas} / 10^{4} ~ \rm{K}$ and $U=|E_{1s}/k_{\rm B}
T_{\rm gas}|$ as in {\tt RECFAST}
code~\cite{1999ApJ...523L...1S}.
Equation \eqref{beta_e} was represented by
$\beta_e = \alpha_e (2\pi m_\mathrm{e} k_\mathrm{B} T_\mathrm{gas}/h_\mathrm{Pl}^2)^{3/2}
 \exp {(E_{2s}/k_\mathrm{B} T_\mathrm{gas})}$ in the original {\tt RECFAST} code.
However, it was pointed out that {\tt RECFAST}
code might overestimate the ionization rate
under the presence of PMFs by Chluba {\it et al.}~\cite{2015MNRAS.451.2244C}.
Therefore, we adopt Eq.~(\ref{beta_e}) to calculate the ionization rate.

For simplicity, we do not include any ionizing photons from astronomical objects
and also assume that there are no helium and heavier elements.
When we calculate Eqs.~(\ref{eq:th}) and (\ref{eq:ion}),
we take into account the fluctuations of
the hydrogen number density $n_\mathrm{H}$ and the IGM density $\rho_\mathrm{b}$,
which are evaluated by Eq.~(\ref{eq:delta_b}).

\subsection{SZ angular power spectrum}
In the previous section, we discuss the effects of the PMFs on the gas
evolution.  When the PMFs are tangled, the fluctuations of the density,
temperature and ionization fraction are generated.  These fluctuations
can induce the temperature anisotropies of the CMB through the so-called
tSZ effect.  In this subsection, we describe the angular power spectrum
of the CMB temperature due to the tSZ effect~(the tSZ angular power
spectrum).

The strength of the tSZ effect is characterized by
the Compton $y$-parameter on the line-of-sight direction~$\hat n$~\cite{1970Ap&SS...7....3S},
\begin{equation}
y (\hat n)\equiv \cfrac{k_\mathrm{B}\sigma_\mathrm{T}}{m_\mathrm{e} c^2}\int {d\chi}~{a_\chi}
w(\chi,\hat n ),
\label{eq:y}
\end{equation}
where $\chi$ is the comoving distance and $a_\chi$ is the scale factor
corresponding to $\chi$.  In Eq.~\eqref{eq:y}, $w(\chi,\hat n )$ is
the function of $n_\mathrm{b}$, $x_{\mathrm{i}}$ and $T_{\rm gas}$ at a comoving
three-dimensional position $\bf x$ given by ${\bf x} = \chi \hat n$,
\begin{equation}
w(\chi, \hat n) = \left. x_{\mathrm{i}} n_\mathrm{b} (T_\mathrm{gas}-T_\gamma)\right|_{\bm{\mathbf x} }.
\label{eq:def_w}
\end{equation}

The CMB temperature anisotropies caused by the tSZ effect can be written
with the Compton $y$-parameter,
\begin{equation}
\frac{\Delta T}{T}(\hat n) = g_\nu y(\hat n),
\label{eq:sz_temp}
\end{equation}
where $g_\nu$ is the spectral function of the tSZ effect,
$g_{\nu} = -4+x/\tanh(x/2)$ with $x\equiv h_{\rm Pl} \nu /k_{\rm B} T$,
and $g_\nu =-2$ in the Rayleigh-Jeans limit of a frequency $\nu$.

According to Eq.~\eqref{eq:sz_temp}, we can obtain the tSZ angular
power spectrum as
\begin{equation}
C_\ell =  \left(\cfrac{g_\nu k_\mathrm{B}\sigma_\mathrm{T}}{m_\mathrm{e} c^2}\right)^2
\int d\chi \cfrac{P_{w}(\chi,\ell/\chi)}{\chi^2},
\label{eq:p_y}
\end{equation}
where $\ell$ is a multipole moment.
Here we adopt Limber's approximation because we are interested in large $\ell$ modes.
In Eq.~\eqref{eq:p_y}, $P_w(\chi, k)$ is the three-dimensional power
spectrum of the Compton $y$-parameter at a comoving distance $\chi$ and
we can obtain $P_w(\chi,k)$ from $w$ in Eq.~\eqref{eq:def_w}.

\section{Simulation Setup}
We can calculate the tSZ angular power spectrum from Eqs.~\eqref{eq:y} and~\eqref{eq:p_y}
by solving Eqs.~\eqref{eq:delta_b},~\eqref{eq:th}~and~\eqref{eq:ion} for a given realization of PMFs.
However, since the source terms due to PMFs for density fluctuations, i.e., Eq.~(\ref{eq:source_b})
and for thermal history, i.e., Eq.~(\ref{eq:source_t}),
are highly nonlinear, it is difficult to obtain these fluctuations in an analytical way.
Therefore we perform numerical simulations
to evaluate the fluctuations of the density, temperature and ionization fraction
and compute the tSZ angular power spectrum from them.

First we assume that the evolution of the PMFs
depends only on the cosmic expansion after the recombination epoch.
That is, we neglect any backreaction from local matter evolution to the PMFs,
and we assume conductivity of the Universe is infinity.
Under these assumptions, the PMFs adiabatically decay with the expansion of the Universe as
\begin{equation}
\mathbf{B}(t,\mathbf{x}) = \cfrac{\mathbf{B}_0(\mathbf{x})}{a^2},
\end{equation}
where $|\mathbf{B}_0|$ is the comoving strength of the PMFs.
Additionally the PMFs are assumed to be statistically homogeneous and isotropic.
Such magnetic fields can be characterized by the power spectrum $P_B(k)$ defined as~\cite{1975mit..bookR....M}
\begin{eqnarray}
\langle B_i^\ast(\mathbf{k}) B_j(\mathbf{k'}) \rangle 
&=& \cfrac{(2\pi)^3}{2} \delta(\mathbf{k-k'})
 \left(\delta_{ij} - \hat k_i \hat k_j \right) P_B(k), \nonumber \\
\label{eq:BB}
\end{eqnarray}
where $\mathbf{B}(\mathbf{k})$ is the Fourier component of
$\mathbf{B}_0(\mathbf{x})$.
For simplicity, we assume a power-law shape for $P_B(k)$ with the spectral
index $n_B$ as
\begin{eqnarray}
P_B(k) &=& \cfrac{n_B+3}{2} \cfrac{(2\pi)^2 B_{n}^2}{k_{n}^{n_B+3}} k^{n_B},
\label{eq:P_B}
\end{eqnarray}
where $B_n$ is the field strength at the normalized scale $k_n$.
We take $k_n$ corresponding to 1 Mpc in this paper.
In this work, we do not consider any specific origin of the PMFs.
Instead, we adopt a general form of power spectrum of PMFs to cover a wide range of PMF models.%
\footnote{%
There are some different definitions of the power spectrum of PMFs,
for example, see Eq.~(3.8) in~\cite{2011PhRvD..84f3010G}.
We follow the above one with which most of previous studies have worked.}
It is useful to introduce $B_\lambda$ which represents the typical
 magnetic field strength on a scale $\lambda$ in the real space as
\begin{equation}
B^2_{\lambda} = \cfrac{1}{2\pi^2} \int^{k_\lambda }_0 k^2 dk P_B(k)
= B_n^2 \left(\cfrac{k_\lambda}{k_n}\right)^{n_B+3},
\end{equation}
where $k_\lambda = 2 \pi/\lambda$.

Additionally, we assume the presence of cutoff scale with PMFs due to
the damping of Alfv\'en waves in the early universe
\cite{1998PhRvD..57.3264J,1998PhRvD..58h3502S}.
This cutoff wave number, $k_c$, is obtained by
\begin{equation}
k_c^{-2} = \cfrac{B_{\lambda_c}^2(t_r)}{4\pi \rho_\gamma(t_r)} \int^{t_r}_0\cfrac{l_\gamma(t')}{a^2(t')} dt',
\label{eq:cut}
\end{equation}
where $t_r$ is the recombination time,
$l_\gamma$ is the mean-free path of CMB photons and $\lambda_c = 2 \pi/k_c$.
In $k \le k_c$, the amplitude of PMFs is derived from Eq.~(\ref{eq:P_B}),
and in $k \ge k_c$, PMFs sharply drop off.

\begin{table}[tb]
\caption{\label{tb:models}The models of PMFs.}
\begin{ruledtabular}
\begin{tabular}{lcdc}
Model &  $B_n$ [nG] & \multicolumn{1}{c}{$n_B$} & $\lambda_c$ [kpc] \\
\colrule
1 & $0.5$ & 0.0 & 250 \\
2 & $0.5$ & -1.0 & 162 \\
3 & $0.1$ & 0.0 & 131 \\
4 & $0.1$ & -1.0 & 72.4
\end{tabular}
\end{ruledtabular}
\end{table}

Resultantly, the statistical character of the PMFs in our model
is determined by only two parameters, $B_n$ and $n_B$.
We investigate the tSZ effects from the PMFs
with some combinations of these parameters summarized in Table \ref{tb:models}.
These parameters are consistent with the Planck constraint on the PMFs~\cite{2016A&A...594A..19P}.

Now we describe the details of our simulations.
First, we set the simulation box size to $(1~\rm{Mpc})^3$.
It is required to resolve the cutoff scale in Eq.~(\ref{eq:cut})
in order to evaluate properly the magnetic field effects.
We adopt the grid number to $64^3$ for models 1--3 and $128^3$ for model 4.
These resolutions are enough to resolve the cutoff scale of magnetic fields.

To satisfy the divergenceless condition of PMFs,
we first make a realization of a vector potential field, $\mathbf{A(k)}$, in wave-number space.
Then, the PMFs, $\mathbf{B(k)}$, are obtained as
\begin{equation}
\mathbf{B(k)} = i \mathbf{k \times A(k)}.
\end{equation}
In this way, the condition $\mathbf{\nabla \cdot B} = 0$ is automatically satisfied.
Also calculating the outer product of $i\mathbf{k}$ and $\mathbf{B(k)}$
provides $\mathbf{\nabla \times B}$ in wave-number space.
Then, we perform the inverse Fourier transformation of these values
to obtain $\mathbf{B}$ and $\mathbf{\nabla \times B}$ in real space.
This procedure allows us to evaluate
the source terms of the density fluctuation and gas temperature
from Eqs.~\eqref{eq:delta_b} and \eqref{eq:th} in real space.
The evolution of physical quantities in all cells of our simulations
are independently calculated from local values,
according to Eqs.~(\ref{eq:delta_b}), (\ref{eq:th}), and (\ref{eq:ion}).
We adopt the fourth-order Runge-Kutta method to solve Eqs.~(\ref{eq:th}) and (\ref{eq:ion}),
and make output data at 67 redshift slices taken logarithmically from $z=1000$ to $z=10$.
We then calculate the power spectrum~$P_w(\chi(z),k)$, at each redshift slice.
We integrate these power spectra with the linear interpolation between the slices and,
finally, we obtain the angular power spectrum in Eq.~\eqref{eq:p_y}.

\section{Results and DISCUSSION} \label{sec:cl}
\begin{figure*}[t]
\begin{tabular}{ccc}
\hspace{-112pt}
\begin{minipage}[m]{0.3\textwidth}
{\includegraphics[width=1.05\textwidth,angle=270]{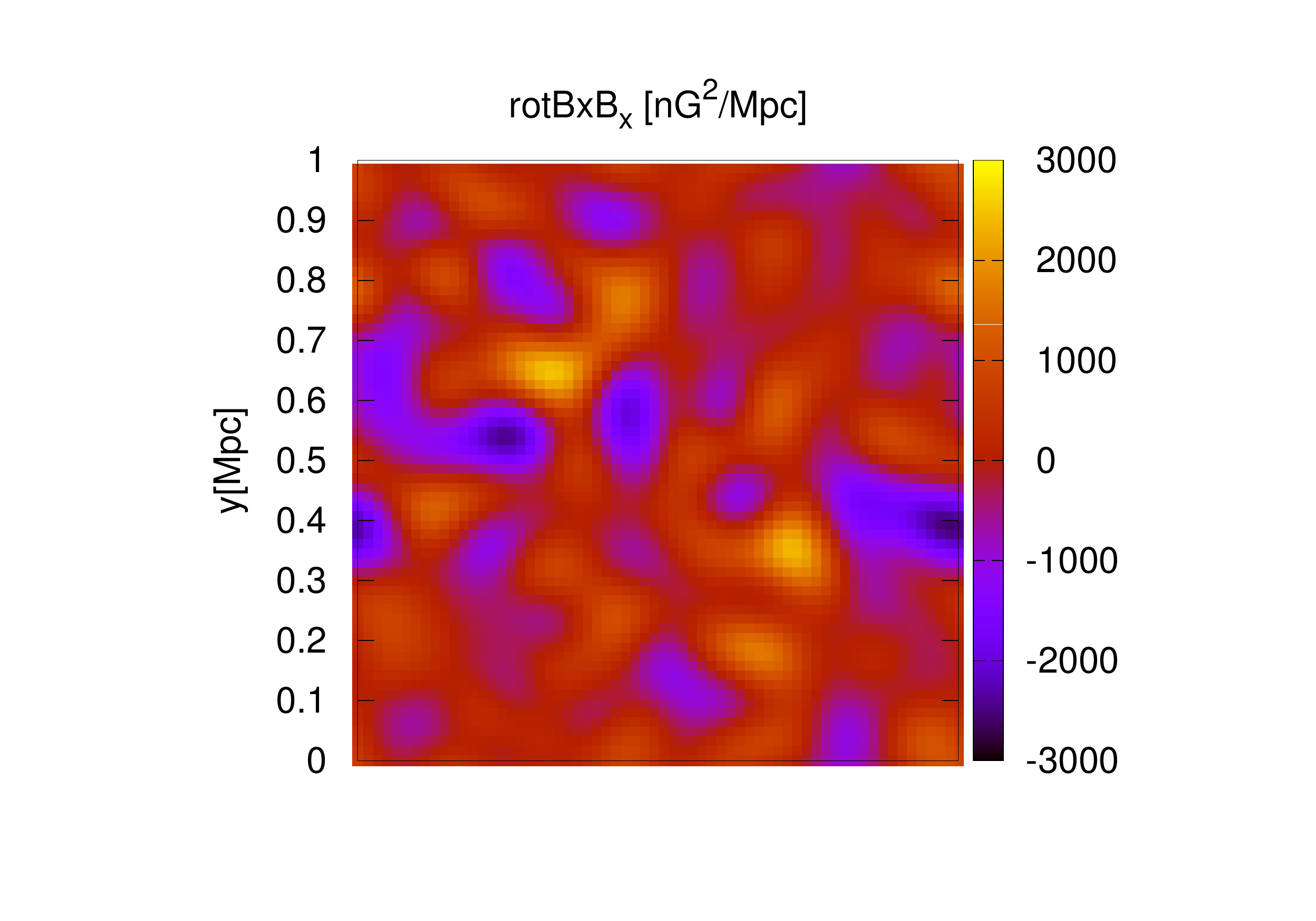}}
\end{minipage}
\hspace{-20pt}
\begin{minipage}[m]{0.3\textwidth}
{\includegraphics[width=1.05\textwidth,angle=270]{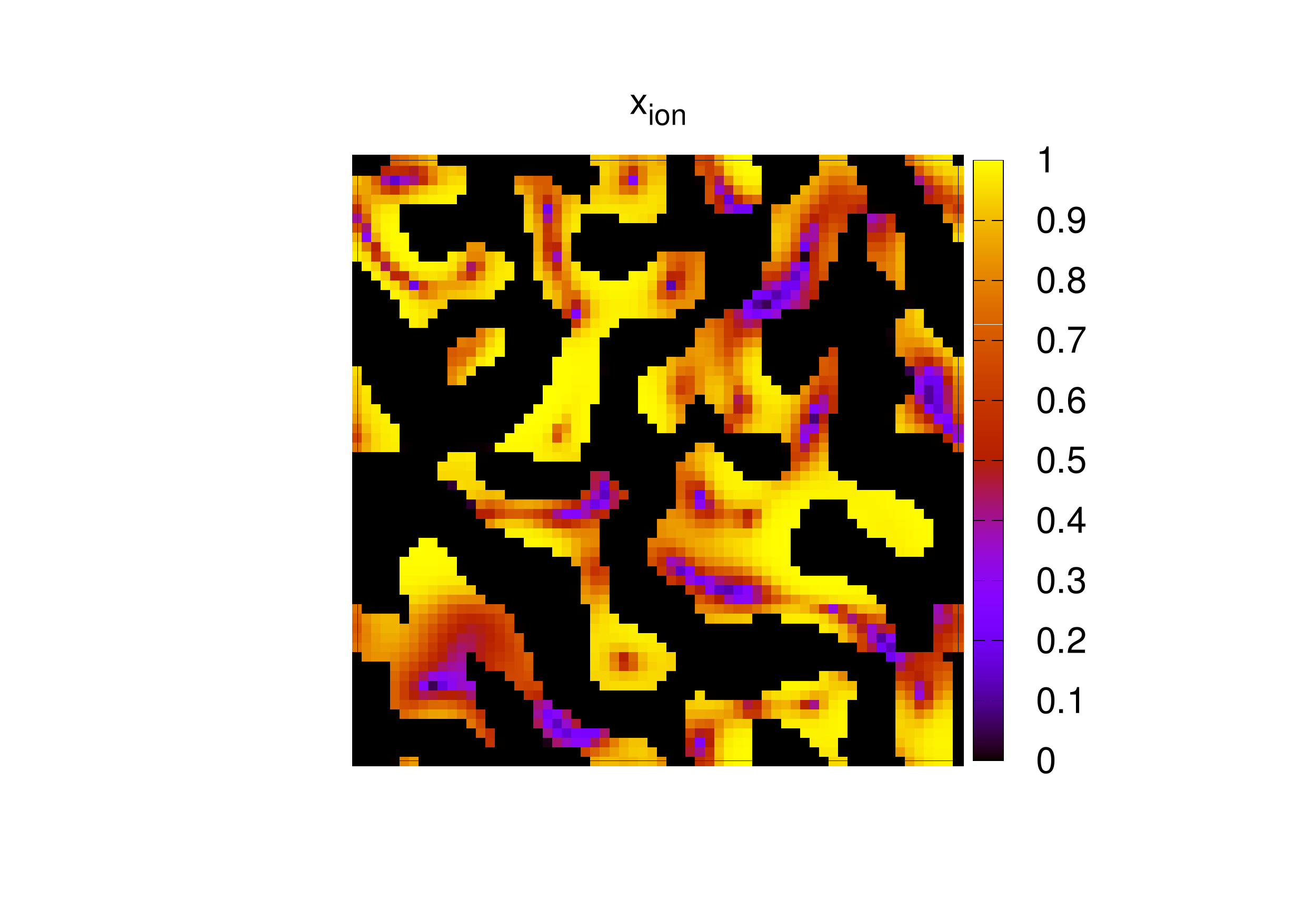}}
\end{minipage}
\hspace{-20pt}
\begin{minipage}[m]{0.3\textwidth}
{\includegraphics[width=1.05\textwidth,angle=270]{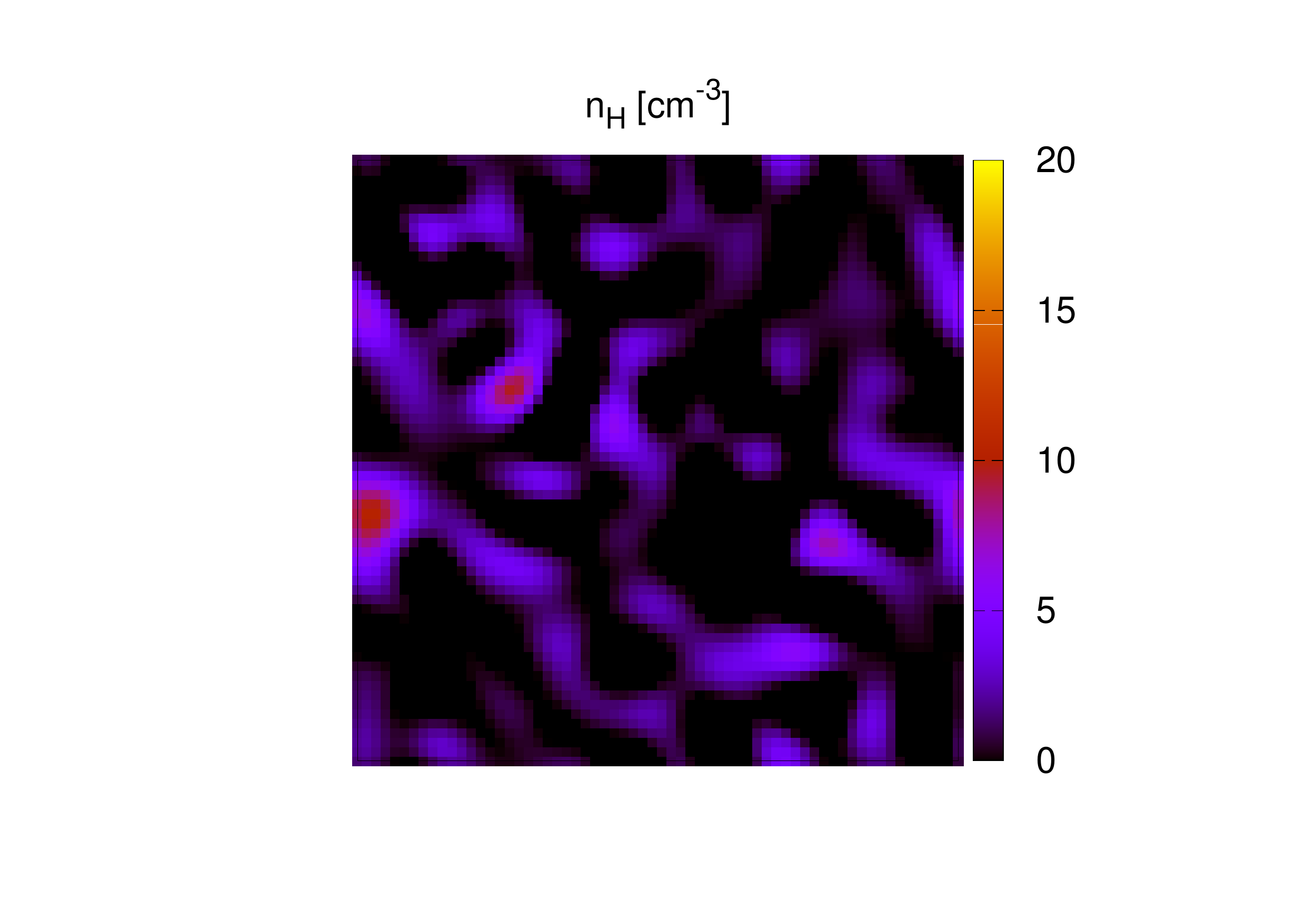}}
\end{minipage}
\vspace{-40pt}
\\
\hspace{-100pt}
\begin{minipage}[m]{0.3\textwidth}
{\includegraphics[width=1.0\textwidth,angle=270]{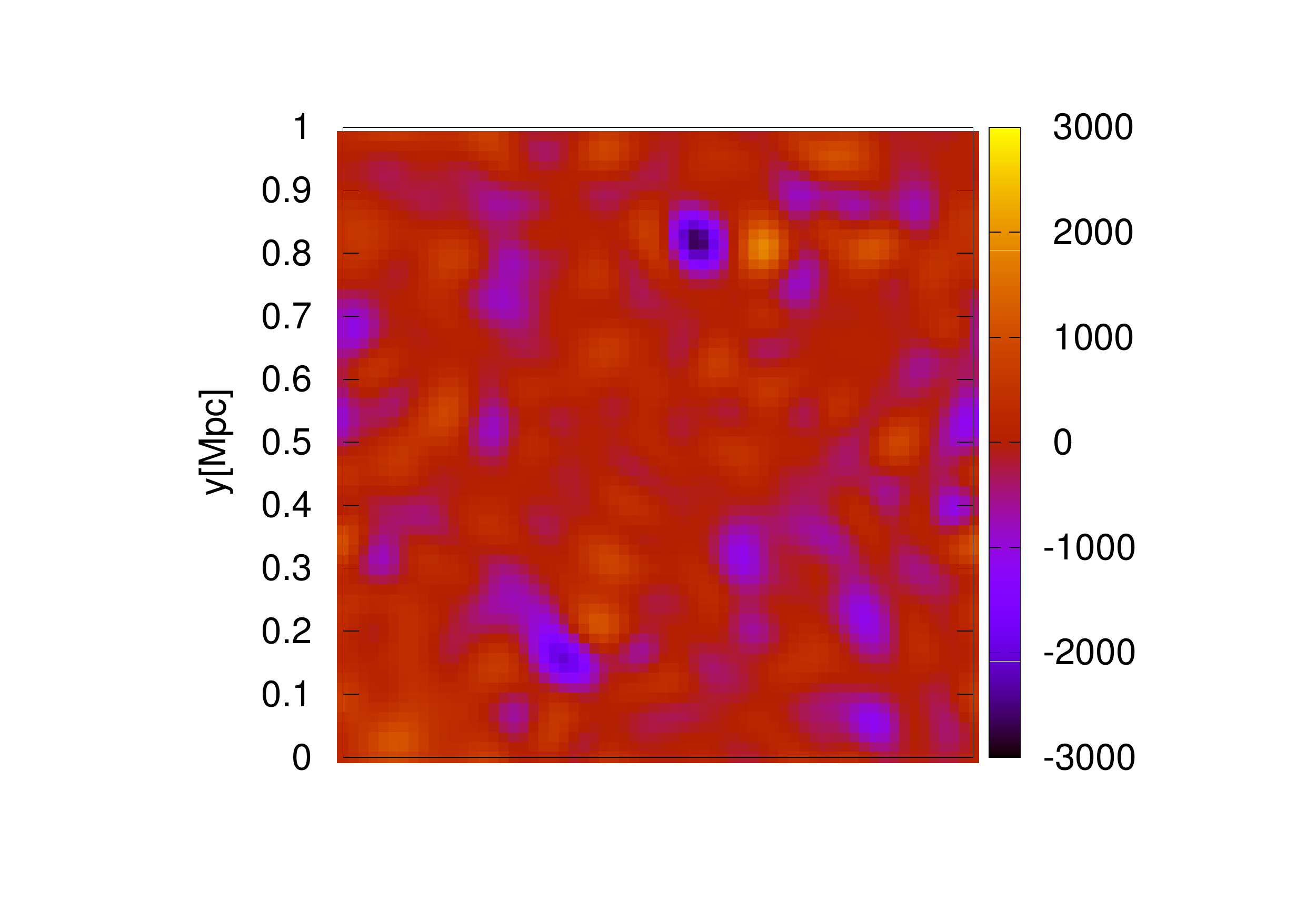}}
\end{minipage}
\hspace{-20pt}
\begin{minipage}[m]{0.3\textwidth}
{\includegraphics[width=1.0\textwidth,angle=270]{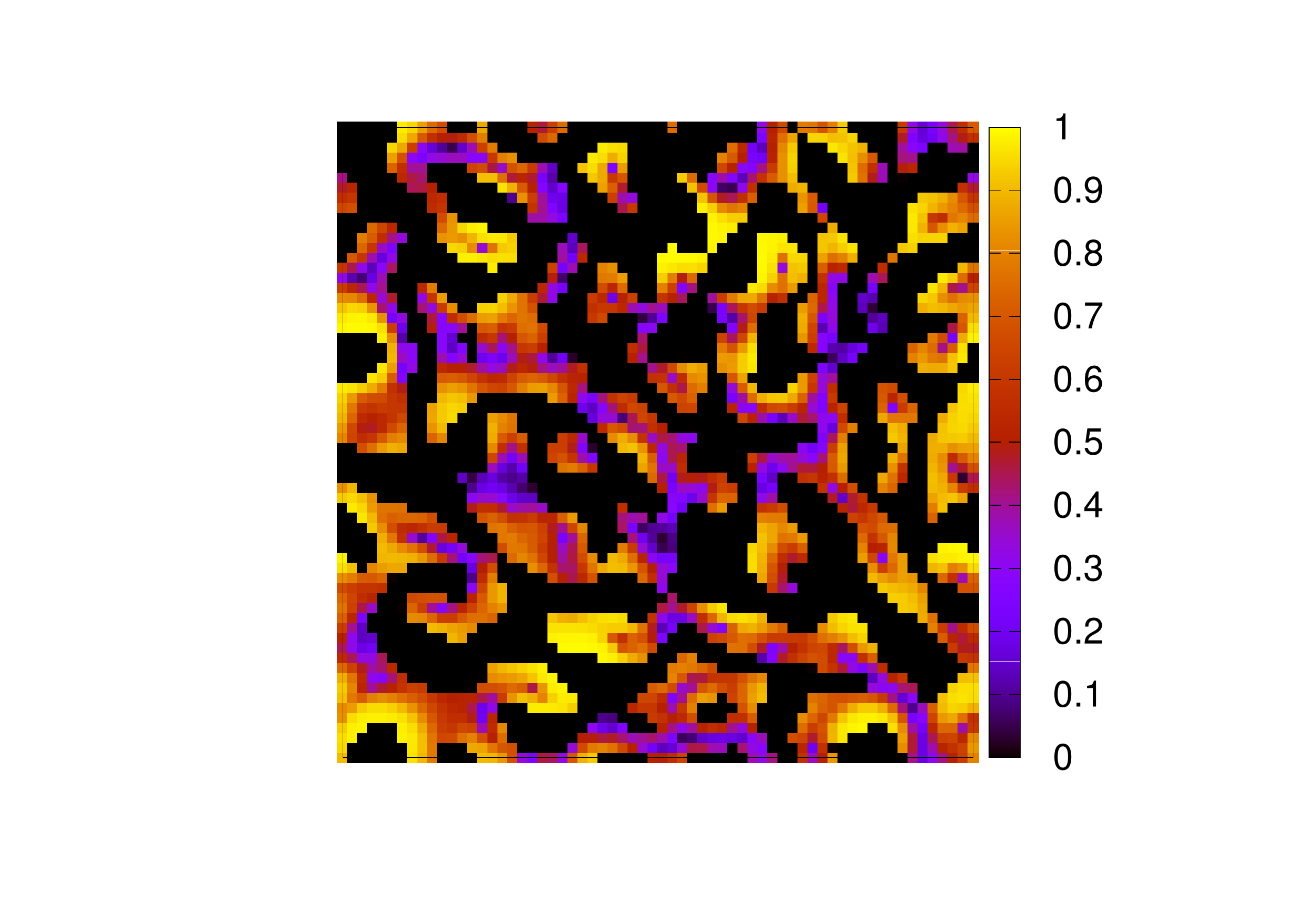}}
\end{minipage}
\hspace{-20pt}
\begin{minipage}[m]{0.3\textwidth}
{\includegraphics[width=1.0\textwidth,angle=270]{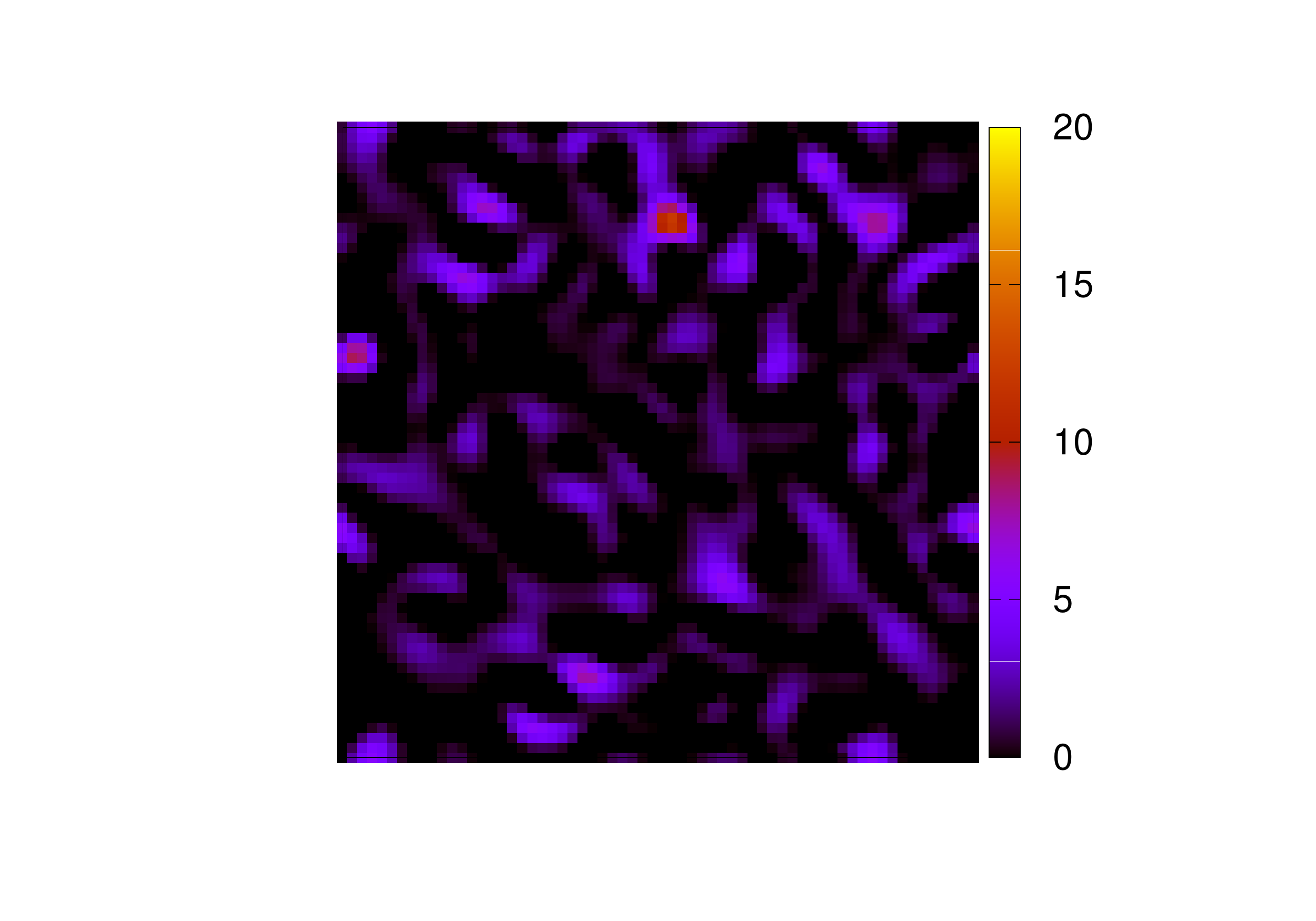}}
\end{minipage}
\vspace{-40pt}
\\
\hspace{-100pt}
\begin{minipage}[m]{0.3\textwidth}
{\includegraphics[width=1.0\textwidth,angle=270]{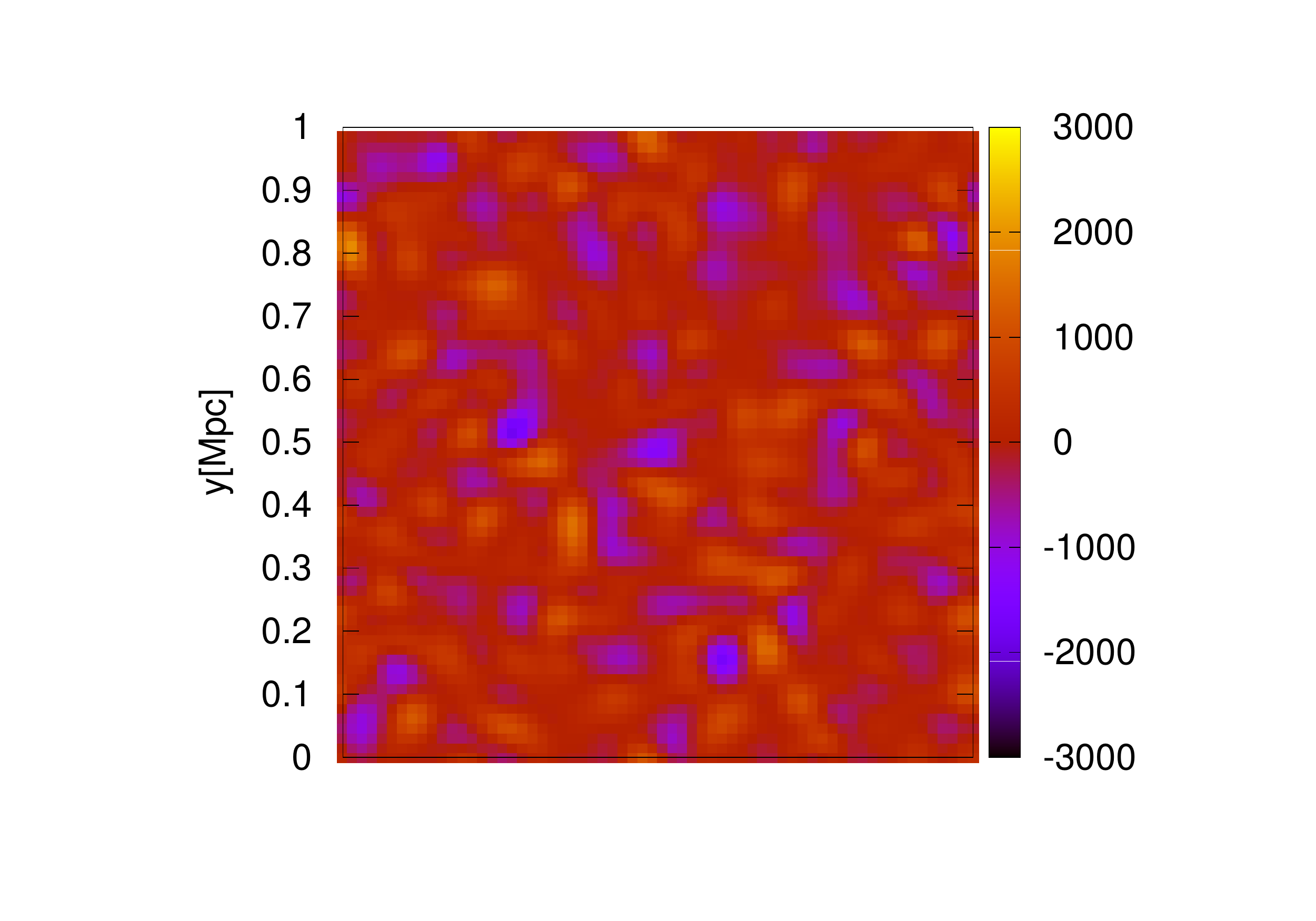}}
\end{minipage}
\hspace{-20pt}
\begin{minipage}[m]{0.3\textwidth}
{\includegraphics[width=1.0\textwidth,angle=270]{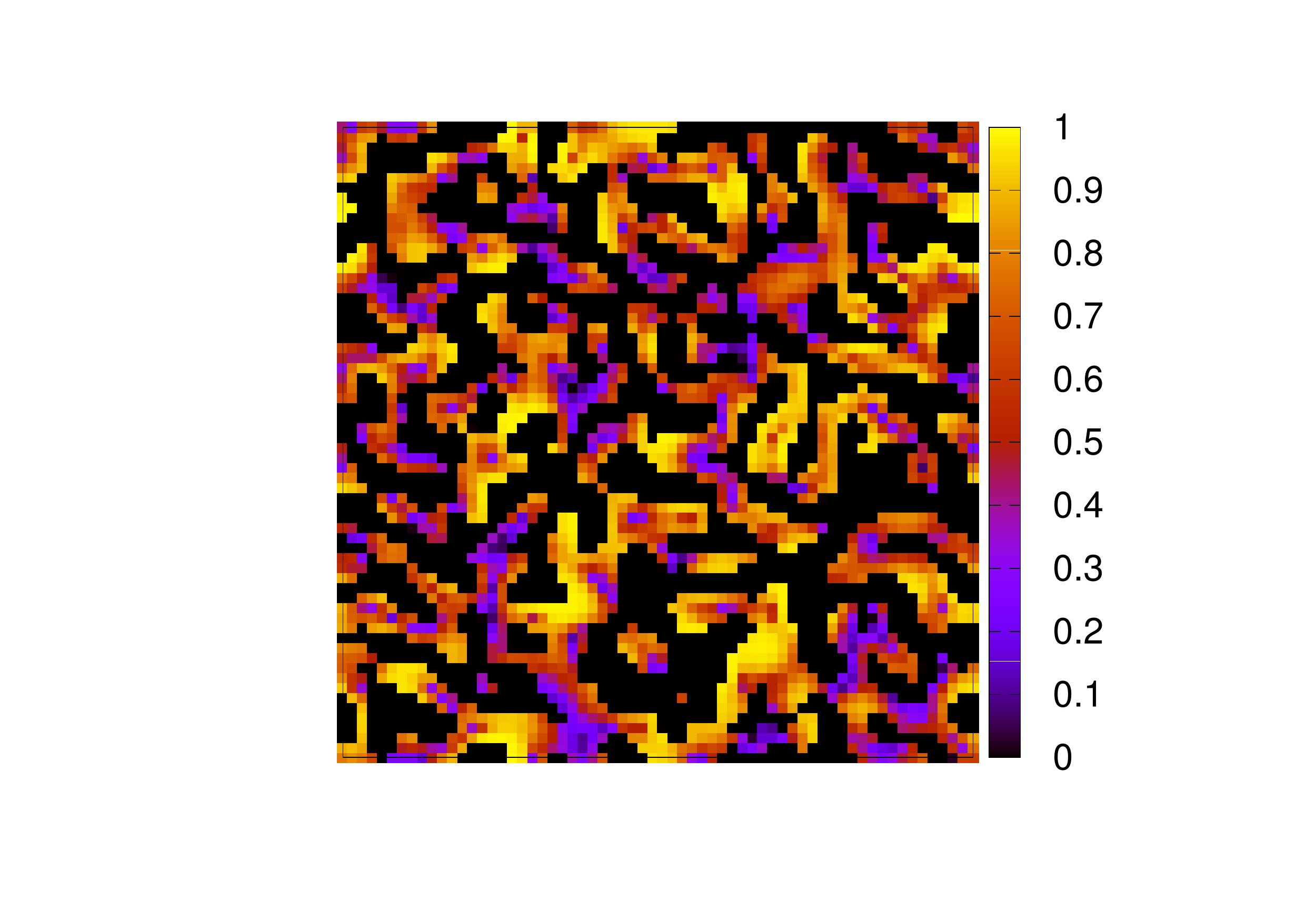}}
\end{minipage}
\hspace{-20pt}
\begin{minipage}[m]{0.3\textwidth}
{\includegraphics[width=1.0\textwidth,angle=270]{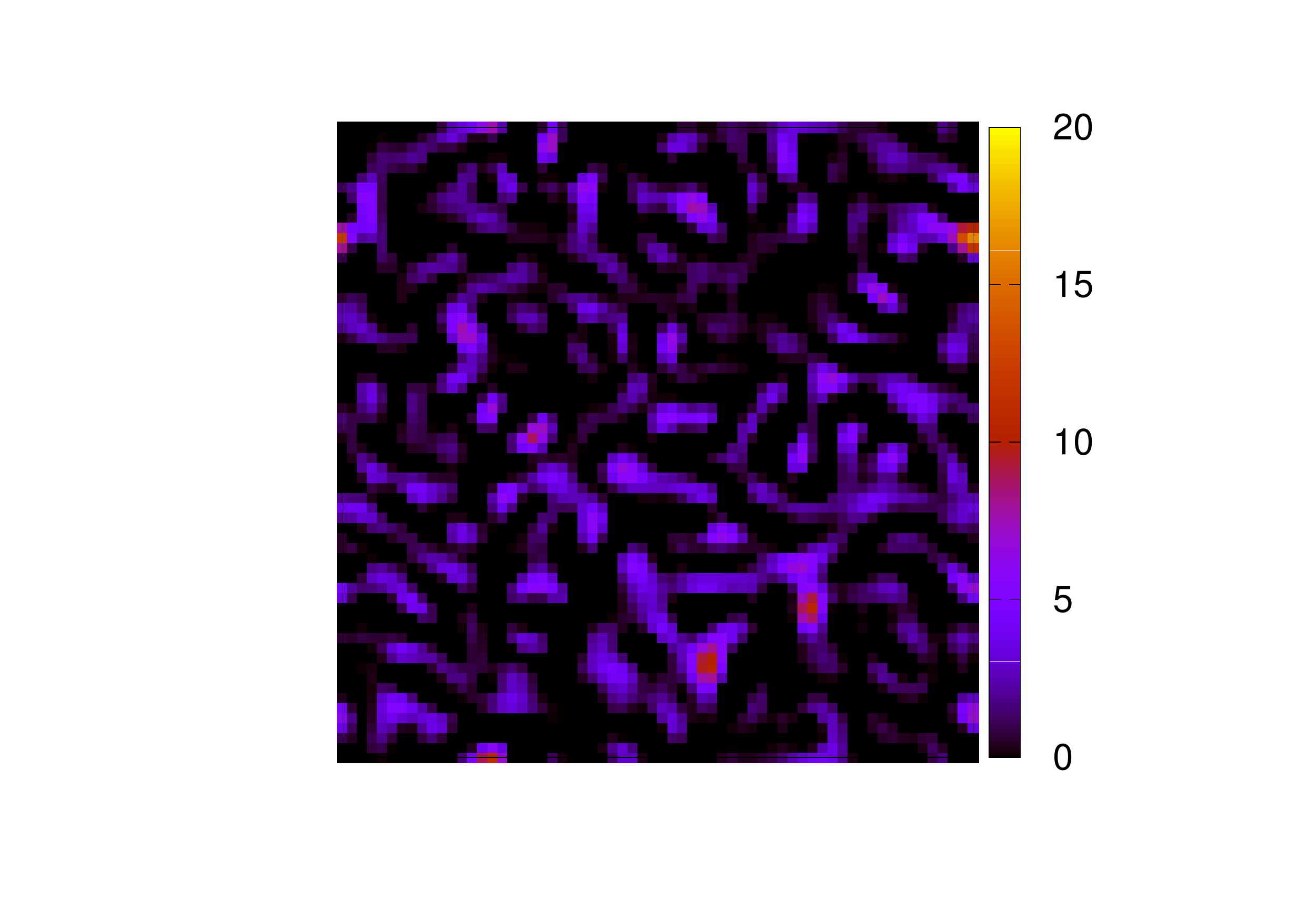}}
\end{minipage}
\vspace{-40pt}
\\
\hspace{-105pt}
\begin{minipage}[m]{0.3\textwidth}
{\includegraphics[width=1.01\textwidth,angle=270]{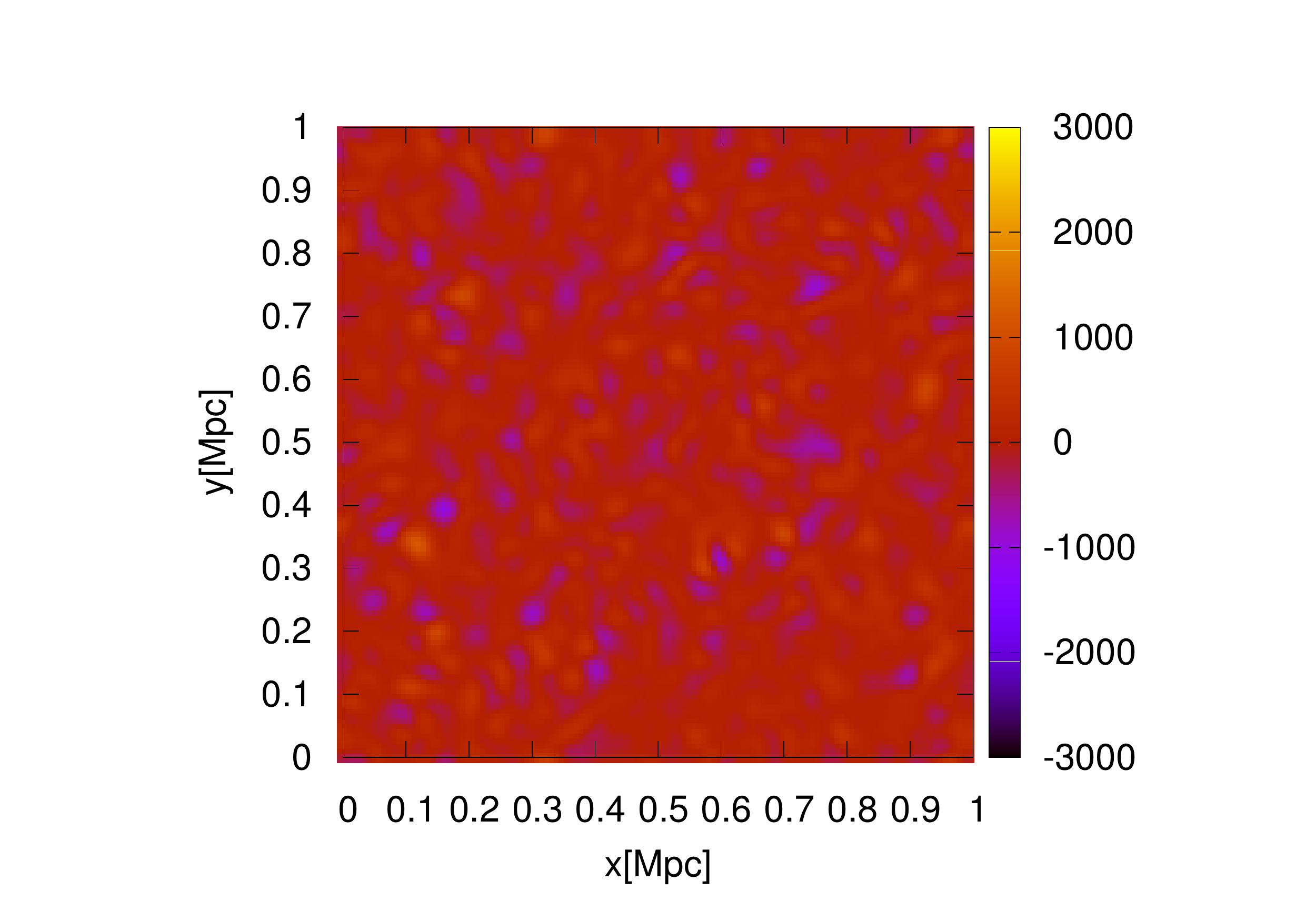}}
\end{minipage}
\hspace{-20pt}
\begin{minipage}[m]{0.3\textwidth}
{\includegraphics[width=1.01\textwidth,angle=270]{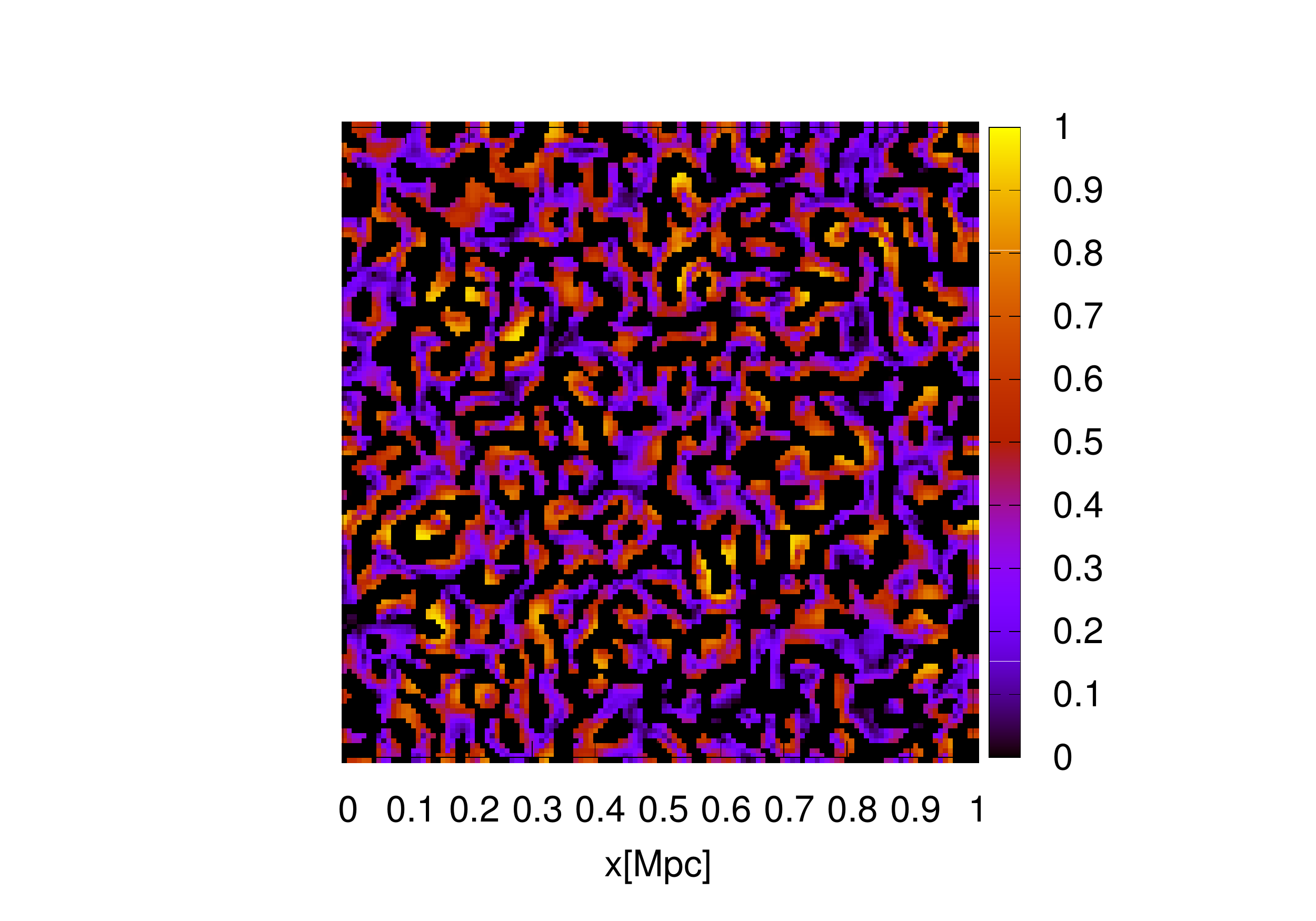}}
\end{minipage}
\hspace{-20pt}
\begin{minipage}[m]{0.3\textwidth}
{\includegraphics[width=1.01\textwidth,angle=270]{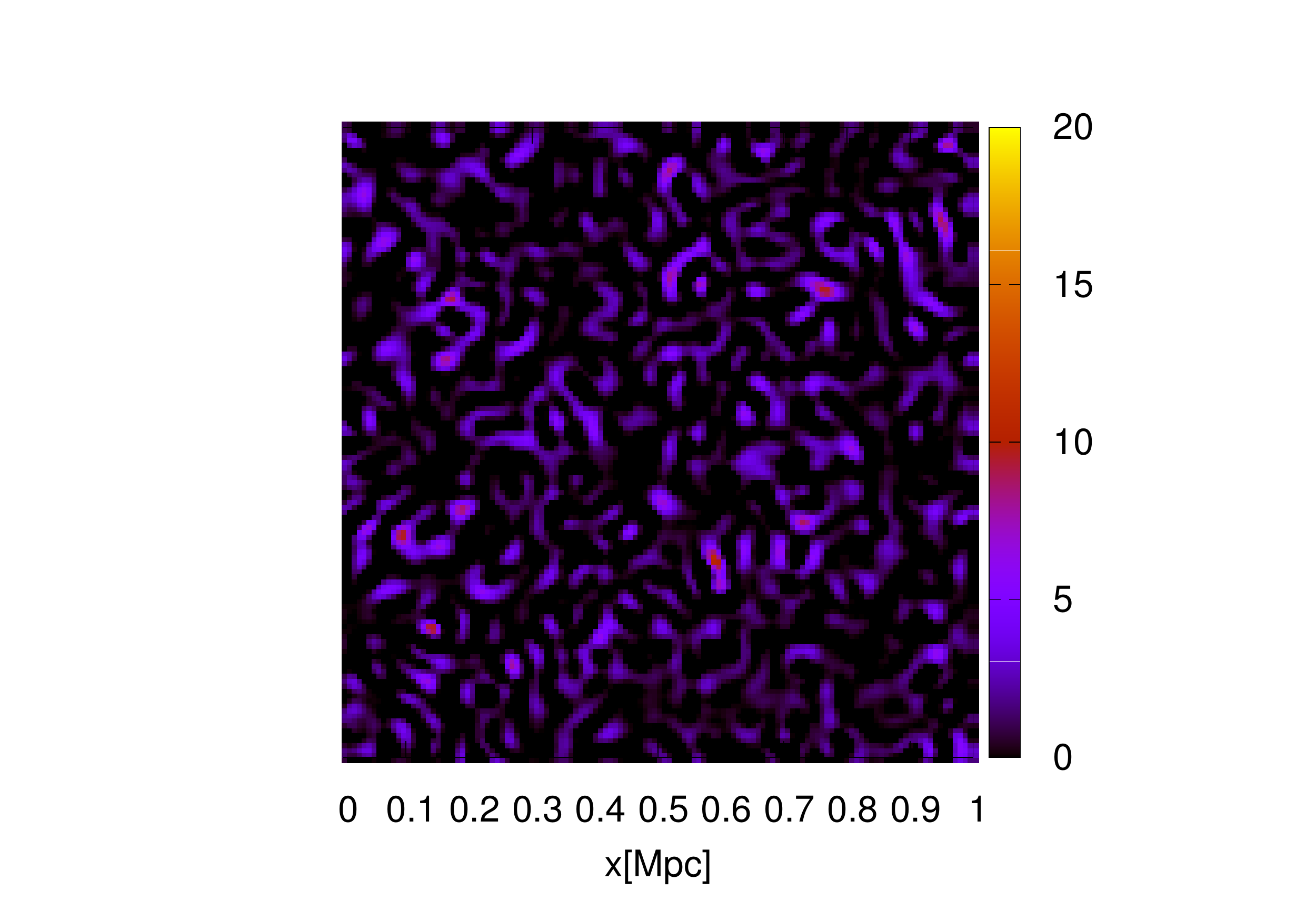}}
\end{minipage}\\
\end{tabular}
\caption{Illustration of $\mathbf{(\nabla \times B) \times B}_x$~(left column),
ionization fraction~(middle column) and the IGM number density~(right column)
in the PMFs models 1--4 from top to bottom at $z=10.0$.
Negative spatial correlation between $x_{\mathrm{i}}$ and $n_\mathrm{H}$ is apparent.
It is also visible that the PMFs with smaller values of $B_{\lambda}$ and $n_B$
generate more homogeneous structure of the Lorentz force and the gas density distributions.} \label{fig:map}
\end{figure*}

We perform our simulations for four different PMF models
with the parameter sets listed in Table \ref{tb:models}.
Depending on the PMFs, the evolutions of the gas quantities,
$\delta_\mathrm{b}$, $T_{\rm gas}$ and $x_{\mathrm{i}}$, are different.
First, we focus on the PMF dependence of these values.

In the left column of Fig.~\ref{fig:map},
we show the two-dimensional structure of the x-component of $\mathbf{(\nabla \times B) \times B}$,
which appears in the source terms due to the Lorentz force
in Eqs.~\eqref{eq:source_b} and~\eqref{eq:source_t}.
In this figure, the panels from top to bottom
correspond to the cases for models 1--4 listed in Table~\ref{tb:models}.
The middle and right columns show
the ionization fraction and the hydrogen number density maps at $z=10.0$, respectively.
In regions where the magnetic fields are strong,
the gas is heated up to $\sim 20000$ K via the ambipolar diffusion.
Furthermore, the gas densities in these regions significantly decrease due to the Lorentz force. 
Since the photoionization becomes more effective than the collisional recombination,
the gas in low density regions maintains a high ionization state.
Thus, it is obvious that $x_\mathrm{ion}$ and $n_\mathrm{H}$ have a strong anticorrelation.
According to Eq.~\eqref{eq:cut}, the cutoff scale becomes small from model 1 to 4 in Table \ref{tb:models}.
Therefore, the model whose cutoff scale is smaller has small-scale structures
in the spatial distribution of $x_\mathrm{ion}$ and $n_\mathrm{H}$.
Note that we set the lower limit of the density fluctuations
to $\delta_\mathrm{b}=-0.9$ to avoid the negative density
because Eq.~\eqref{eq:delta_b} evolves the density fluctuations to $|\delta_\mathrm{b}|\gg 1$ quickly,
in particular, on small scales.
Since it is a linearized, Eq.~\eqref{eq:delta_b} is not valid in such a highly nonlinear regime.
We discuss this point in the end of this section.

\begin{figure*}[t]
\hspace{-3cm}
\begin{minipage}[m]{0.3\textwidth}
{\includegraphics[width=1.0\textwidth,angle=270]{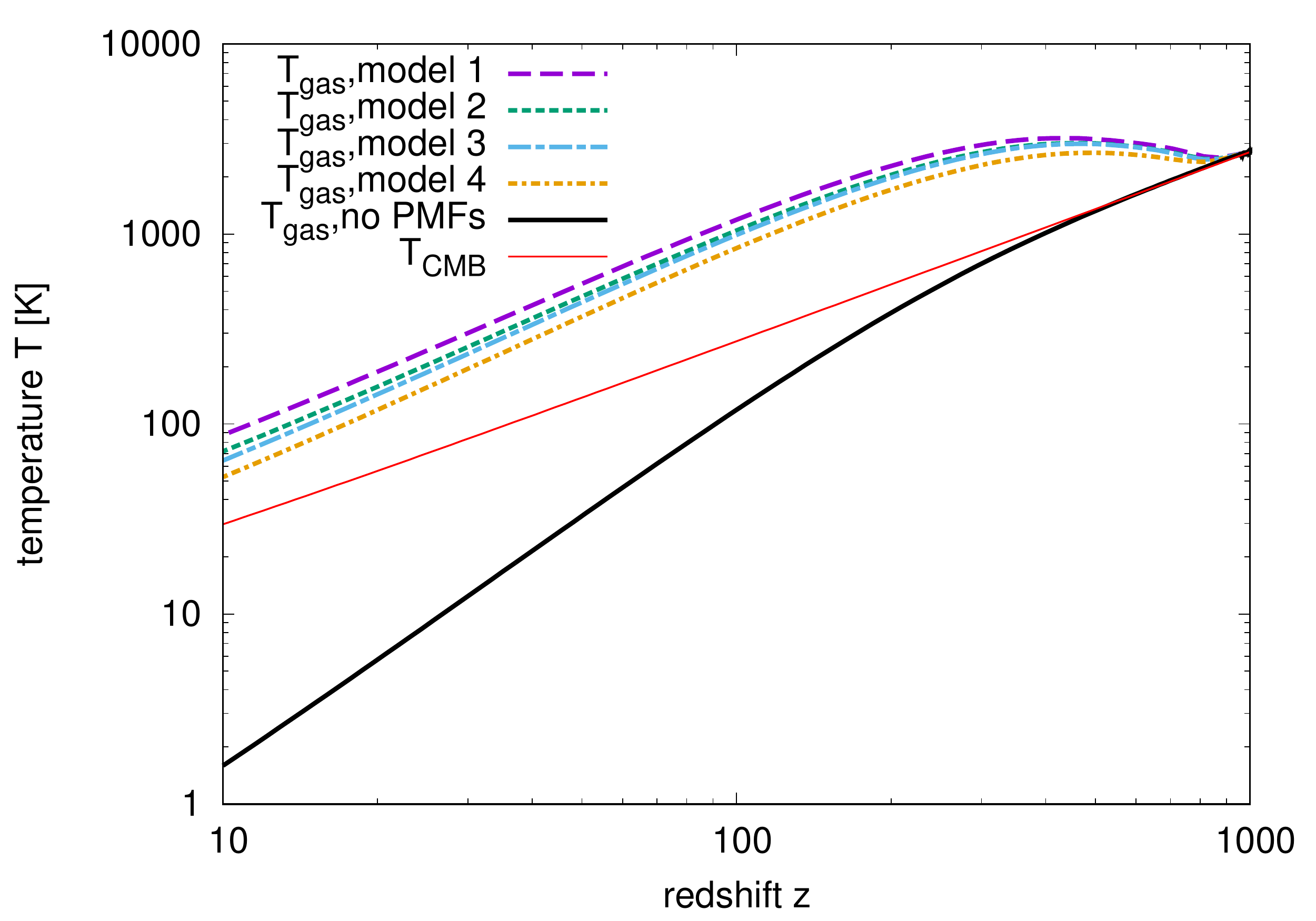}}
\end{minipage}
\hspace{2cm}
\begin{minipage}[m]{0.285\textwidth}
{\includegraphics[width=1.0\textwidth,angle=270]{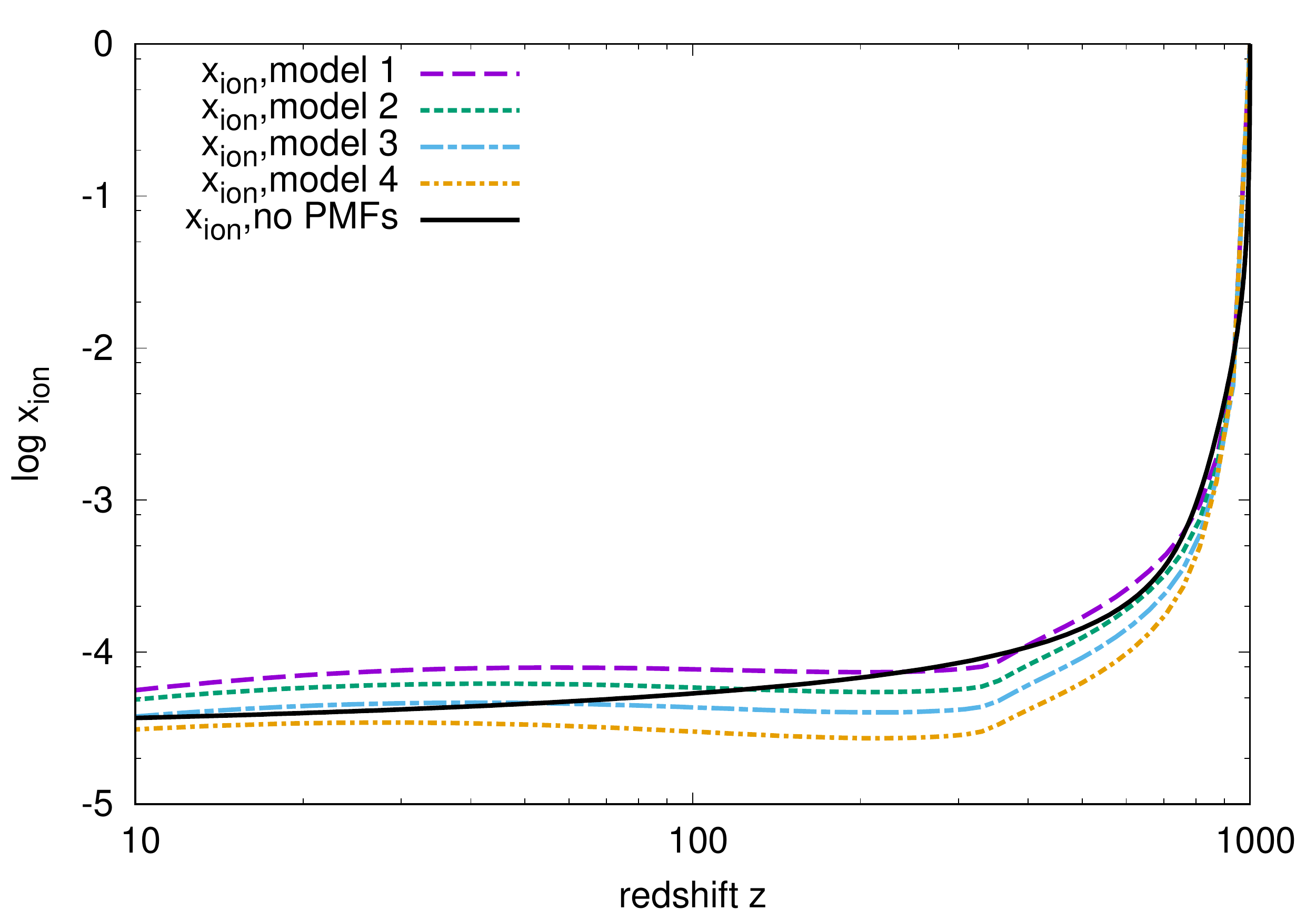}}
\end{minipage}
\vspace{1.5mm}
\caption{
Time evolutions of the gas temperature~(left) and ionization fraction~(right)
with and without the PMFs.
The lines are the mass weighted averages of these values
in the corresponding models shown in the legend.
The black wide solid lines are time evolutions without the PMFs,
and in the left panel the CMB temperature is also shown by the red line.
If the PMFs exist, the gas temperature is always larger
than that without the PMFs due to the ambipolar diffusion.
On the other hand, the values of the ionization fraction with PMFs
tend to be  smaller than that without PMFs at high redshifts,
and start to increase after $z\sim300$ due to the enhanced recombination rate
at high density regions induced by the PMFs. 
See the main text for details.
}
\label{fig:mean_rho}
\end{figure*}

Fig.~\ref{fig:mean_rho}
shows the time evolutions of the average gas temperature and the ionization fraction
for different models of the PMFs.
These values are obtained by density-weighted averaging.
Therefore Fig.~\ref{fig:mean_rho} mainly reflects the evolution in highly overdense regions.
The PMF heating term in Eq.~\eqref{eq:th} decreases with increasing the IGM gas density.
Therefore the heating becomes less effective as the gas becomes denser.
Indeed, the left-hand side of Fig.~\ref{fig:mean_rho} shows that
the gas temperature is saturated at 3000--4000 K
around $z\sim300$.
In such high redshifts, the density contrast has not yet grown very much,
and, as a result, the heating efficiency is still high.
However, the gas temperature starts to drop after $z\sim300$
because the density contrast develops well and it decreases the heating efficiency.
Fig.~\ref{fig:mean_rho} also shows that
the dependence of the gas temperature on the PMF models is small.
Even at $z=10$, the difference of the gas temperature between 0.5 and 0.1 nG is within the factor of 2.
This is because the saturated temperature in high redshifts is independent on the PMF model
and depends mainly on the several cooling effects and the ionization fraction.
After the saturation, the temperature slowly goes down
balancing between the cooling and heating effects.
As a result, the PMF dependence of the temperature
is small for the magnetic field strengths considered in this work.
Although it is not apparent in the figure as mentioned above,
we investigate the gas temperature in the low density regions.
We find that the gas temperature in such regions increases soon after the recombination epoch similarly.
After reaching the temperature saturation,
however, the gas temperature keeps as high as the saturated temperature even in the lower redshifts.

The right-hand side of Fig.~\ref{fig:mean_rho} shows
the density-weighted average value of the ionization fraction for different PMF models.
Contrary to the gas temperature evolution,
the average value of the ionization fraction does not simply reflect the values in high density regions.
This is because the typical ratio of the ionization fraction of the high density region to the low density region
is much greater than that of the density contrast,
i.e., $(\rho x_{\mathrm{i}})^{\rm high~density} < (\rho x_{\mathrm{i}})^{\rm low~density}$.
As we can expect,
the ionization fraction in lower density regions keeps $x_i\approx 1$ after the recombination epoch.
This is because the gas density is low and the collisional recombination term in Eq.~\eqref{eq:ion} becomes small.
In high density regions, as the density evolves,
the collisional recombination becomes effective and the ionization fraction quickly drops down.
We find that, while $x_{\mathrm{i}}\approx 1$ at the lowest density regions
where $\delta_\mathrm{b} = -0.9$, $x_{\mathrm{i}} \approx 10^{-7}$
at the highest density regions with $\delta _b > 10^3$.
Because of this huge gap in $x_{\mathrm{i}}$ between low and high density regions,
the density-weighted average value of $x_{\mathrm{i}}$
does not only reflect $x_{\mathrm{i}}$ in high density regions.
Basically, the full ionization in low density regions makes the average value of the ionization fraction
larger than that without PMFs plotted in black in Fig.~\ref{fig:mean_rho}.
On the other hand, the almost neutral gas in high density regions
can decrease the average value below that without PMFs
(see the yellow short dashed curve around $z \sim 600$ in the right panel of Fig.~\ref{fig:mean_rho}).
In summary, the average value of $x_{\mathrm{i}}$ is determined by the balance
between low and high density regions.
Although model 1, in which the strength of magnetic fields at the cutoff scale is strongest,
has the largest value of $x_{\mathrm{i}}$, the model dependence is not so significant.

\begin{center}
\begin{figure}[t]
{\includegraphics[width=0.3 \textwidth, angle=270]{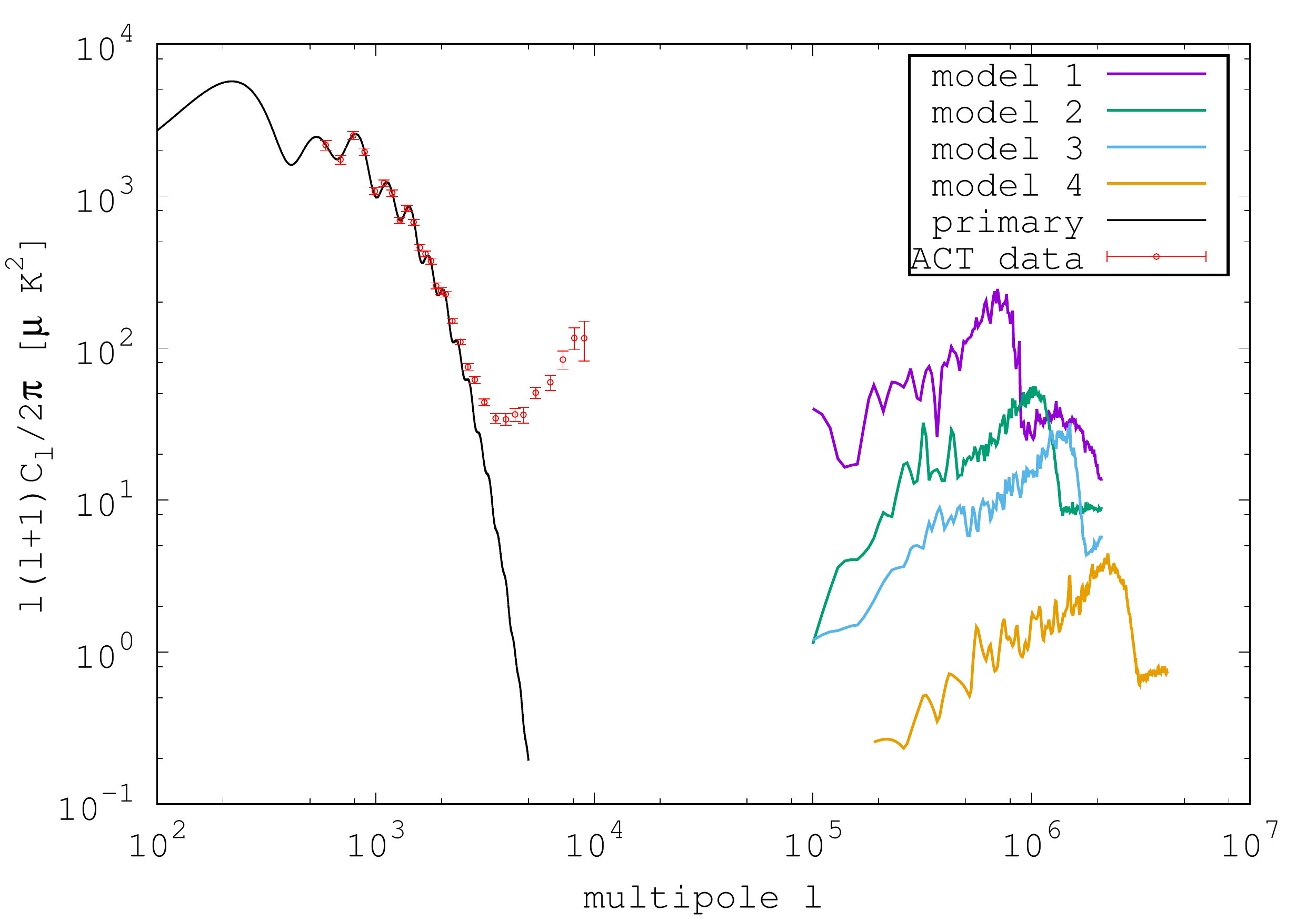}}
\vspace{2mm}
\caption{
Angular power spectra of the tSZ effect induced by the PMFs,
obtained by integrating Eq.(15)  from $z = 1000$ to $z=10$,
for the four models listed in Table \ref{tb:models}.
The primary CMB angular power spectrum
and the observational data with the Atacama Cosmology Telescope (ACT) \cite{2013JCAP...07..008H}
are also shown by the black solid line and the red dots with error bars, respectively.
Clearly, the present observational data cannot constrain the PMF models 1--4
because the angular resolution is not enough.
}
\label{fig:Power}
\end{figure}
\end{center}

We show the CMB temperature angular power spectrum
due to the tSZ effect in the IGM caused by PMFs in Fig.~\ref{fig:Power}.
The tSZ angular power spectrum has a peak around the cutoff scale of the PMFs
and the amplitude depends on the PMF strength at the cutoff scale.
Therefore, the angular spectrum in model 1 has the largest amplitude among our PMF models.
However, it is difficult to provide the dependence of the power spectrum amplitude
on the PMF parameters in the analytical form.
This is because the physical gas quantities related to the tSZ effect
are highly nonlinear and become saturated in some regions.
We also find that the tSZ angular power spectrum decays proportional to $\ell$ on larger scales
than the cutoff scale independently on the spectral index of the PMF,~$n_B$.
This means that the tSZ effect comes from the magnetic field predominantly on the cutoff scale,
and magnetic fields on larger scales have little impact on the tSZ CMB power spectrum.
Although we have shown the results with $n_B =0.0$ and $-1.0$ in Fig.~\ref{fig:Power}, 
we have confirmed that $n_B$ indeed affects tSZ anisotropies only through the cut-off scale
in the cases with $-1.0<n_B<2.0$.
However, it does not necessarily mean that
the tSZ angular power spectrum is insensitive to $n_B$ because the cutoff scale does depend on $n_B$.
Therefore, we conclude that,
although the measurement of the tSZ effect due to the PMFs
can provide the information about the cutoff scale of the PMFs,
it is required to perform careful comparison between the observational data and the theoretical prediction
to deduce the properties of the PMFs, such as the field strength and the spectral index.

At the end of this section,
we make comments on the validity of the gas density evolution
and the impact on the final results.
To obtain the density evolution,
we solve Eqs.~(\ref{eq:cdm}) and (\ref{eq:baryon}) in which we make two important assumptions,
i.e., neglecting the thermal pressure and employing linear perturbations.
For the validity of the former assumption,
we have confirmed that
the grid scales of our simulations are always larger than the Jeans scale.
On the other hand, as already mentioned above,
we find that there are many regions
where the density contrast is much larger than unity
and the linear approximations are no longer valid.
However, such high density regions have a tiny ionization fraction in general
due to the collisional recombination process.
As a result, the contribution to the tSZ angular power spectrum is negligibly small.
The overestimation of the gas density does not harm our final results.
As regards low density regions,
we set the lower limit of the IGM density contrast to $\delta_\mathrm{b} = -0.9$
in order to avoid a negative IGM density.
This procedure means that we artificially take into account the nonlinear structure formation,
that is, the void formation,
because voids are observed as significant underdense regions
with $\delta_\mathrm{b} <-0.85$~\cite{2012MNRAS.421..926P}.
This limit violates the mass conservation in a simulation box.
However, although this violation leads to the overestimation of the density in high density regions,
it does not seem to give a negative impact
on the estimation of the $y$-parameter in low density regions
which produces significant contributions on the tSZ angular power spectrum.
To confirm this point,
it is required to calculate the tSZ spectrum
including the nonlinear effect in the IGM density evolution.
We perform the numerical simulations in our future work to take into account this nonlinear effect.

\section{Conclusion}
In this paper we have investigated the impact of the PMFs on the CMB temperature anisotropies
caused by the tSZ effect in the IGM.
Here we have taken into account two effects of PMFs on the IGM including their spatial correlations;
the generation of the IGM density fluctuations by the Lorentz force,
and the heating of the IGM through the ambipolar diffusion.
The spatial inhomogeneity of PMFs can induce the fluctuations of the IGM density,
temperature and ionization fraction through these effects.
The anisotropy of the Compton $y$-parameter in the IGM arises due to the existence of these fluctuations.
We have calculated the tSZ angular power spectrum,
assuming the PMF statistical properties,
i.e., the amplitude of the power spectrum and the spectral index.

To evaluate the tSZ angular power spectrum,
we have performed numerical simulations of the IGM evolution
with the realization of random Gaussian PMFs.
This is the first attempt to investigate the effect of PMFs
on the spatial distributions of the IGM gas properties consistently.
Performing numerical simulations for different PMF models,
we have found that the scale of the spatial distributions corresponds to the cutoff scale of the PMFs.
There are strong relations among the IGM density, temperature and ionization fractions.
In high density regions, since the heating rate of the PMF per one IGM particle becomes low,
the IGM temperature is not effectively heated and the ionization fraction drops down
because of the enhancement of the recombination rate due to the high density.
On the other hand, in low density regions,
the collisional recombination occurs less effectively than that in high density regions,
and thus the ionization fraction can be still high in the lower redshifts.

From the results of our simulations,
we have calculated the tSZ angular power spectrum due to the IGM with PMFs.
We have found that since high density regions are almost neutral,
their contributions to the tSZ power spectrum are almost negligible.
Therefore the SZ measurement can probe mainly lower density regions heated by the PMFs.
The tSZ angular power spectrum has a peak around the cutoff scale of
the PMFs~($\ell \sim 10^6$ for sub-nG PMFs) and
its amplitude depends on the PMF strength at the cutoff scale.
On such small scales,
the tSZ effect in galaxy clusters and the kSZ effect due to the patchy reionization
can produce the CMB temperature anisotropies.
However these signals have the peaks around $\ell \lesssim 10^4$ and decay on high $\ell$ modes.
On the other hand, the tSZ angular power spectrum even for $0.1$~nano Gauss PMFs
keeps increasing up to $\ell \sim 10^6$.

In this work, we have used the linearized equation to calculate the density evolution of the IGM.
It is known that the density fluctuations generated by the PMFs have the blue spectrum.
Therefore, in our simulations,
there are many regions where the IGM density contrast is greater than unity
and the linearized equation is no longer valid there.
However, as mentioned above,
such high density regions have significantly small ionization fraction and their contributions to the
Compton $y$-parameter are expected to be negligibly small.
That is, the existence of much higher density contrast than unity
does not make us overestimate the tSZ angular power spectrum.
In low density regions, we set the bound of the density contrast,
$\delta_\mathrm{b} > -0.9$ to avoid the negative density.
Imposing this bound,
we intend to take into account
approximately the nonlinear effect of the structure formation in a low density region,
i.e.,~a void formation.
The density evolution is determined by the local strength of the Lorentz force
in our simulation based on the linear density perturbation theory.
However, the formation of voids also depends on the environmental condition.
Therefore, it is required to include the nonlinear effect of the structure formation
to evaluate the tSZ angular power spectrum properly.
Besides, we only consider the cosmological expansion in the PMF evolution.
Even in low ionization fraction, PMFs could be frozen in the IGM
and the density evolution gives the effect on the PMF evolution.
This PMF evolution can affect the thermal history of the IGM, in particular,
in high density regions and may enhance the tSZ angular power spectrum.
To improve these simplified treatments,
the detailed MHD simulation of cosmological structure formation with PMFs is required.
Furthermore, such a simulation allows us to investigate the PMF effect on the collapse condition
\cite{2014JCAP...08..017S}
and the enhancement of the tSZ angular power spectrum due to galaxy clusters
\cite{2011MNRAS.411.1284T,2012PhRvD..86d3510S}.
 
We have shown that the PMFs can generate tSZ signal on small scales after the recombination epoch.
Therefore it is worth mentioning about the possibility
to provide constraints on the PMFs from small-scale CMB observations.
In the current observation status,
the foreground emissions dominate on small scales where the tSZ signal from the PMFs arises,
and it is difficult to remove the foreground.
In addition to the tSZ signal studied in this work,
the existence of PMFs can create non-negligible small-scale CMB anisotropy
before the recombination epoch through the Doppler effect
due to the velocity perturbations induced by the PMFs
~\cite{1998PhRvL..81.3575S, 2006PhRvD..74f3002G}.
Therefore, in order to obtain constraints on the PMFs,
it is required to precisely investigate the CMB anisotropies on small scales
including all of these contributions.
However, it is beyond the scope of this paper and we address this issue in the future.

%%%%%%%%%%%%%%%%%%%%%%%%%%%%%%%%%%%%%%%%%%%%%%%%%%%%%%%%%
\acknowledgments
{We would like to thank the anonymous referee for helpful comments.
We thank S. Saga, S. Asaba, K. Horiguchi, K. Yoshikawa, J. Chluba,
and A. Kusenko for useful discussions about baryon density fluctuations. 
This work is supported by KAKEN Grant-in-Aid for Scientific Research,
(B) No.~25287057~(K.I. and N.S.), (A) No.~17H01110~(N.S.), 16H01543~(K.I.),
and for Young Scientists (B) No.~15K17646~(H.T.).
}

%\appendix*
%\section{appendix}
%%%%%%%%%%%%%%%%%%%%%%%%%%%%%%%%%%%%%%%%%%%%%%%%%%%%%%%%%%%%%%%%%
%

\end{document}